\documentstyle[11pt,newpasp,twoside,psfig]{article}
\index{summary}
\index{instructions}
\index{template}

\def\edcomment#1{\iffalse\marginpar{\raggedright\sl#1\/}\else\relax\fi}
\reversemarginpar

\begin{document}
\title{Chandra reveals the Dynamic Structure of the Inner Crab Nebula}
\author{Koji Mori}
\affil{Department of Astronomy and Astrophysics, 525 Davey Laboratory,
The Pennsylvania State University, University Park, PA 16802, U.S.A.}
\author{J. Jeff Hester}
\affil{Department of Physics and Astronomy, Arizona State University,
Tempe, AZ 85287-1504, U.S.A.}
\author{David N. Burrows, George G. Pavlov}
\affil{Department of Astronomy and Astrophysics, 525 Davey Laboratory,
The Pennsylvania State University, University Park, PA 16802, U.S.A.}
\author{Hiroshi Tsunemi}
\affil{Department of Earth and Space Science, Graduate School of
Science, Osaka University, 1-1 Machikaneyama, Toyonaka, Osaka 560-0043
JAPAN}

\begin{abstract}
We present a series of monitoring observations of the Crab Nebula with
the {\it Chandra X-ray Observatory}, focusing on the temporal
evolution of the structure. This series of 8 observations, spanning a
period of approximately six months, shows the dynamic nature of the
inner X-ray structures. We detected outward moving ``wisps'' from the
recently discovered inner ring seen in optical observations.  We also
find that the inner ring itself shows temporal variations in
structure. The torus also appears to be expanding. Such temporal
variations generally match the canonical scenario that an expanding
synchrotron nebula injected from the pulsar is confined by the
supernova ejecta.
\end{abstract}

\vspace{-0.5cm}
\section{Introduction}
\vspace{-0.2cm}

The Crab Nebula has been the best laboratory for investigating the
mechanism linking the pulsar wind nebula (PWN) with the pulsar. The
energetics confirms that the PWN is powered by the spin-down energy of
the pulsar, and it is generally believed that this energy is
transported by a relativistic wind (e.g. Kennel \& Coroniti 1984).
Hester (1998) and Tanvir, Thomson, \& Tsikarishvili (1997) showed that
the ``wisps'', which are elliptical ripples around the pulsar (Scargle
1969), are moving outwards with a speed of about 0.5{\it c}. Such high
energy phenomena must be associated with X-ray emission. Here, we
present the results of a series of monitoring X-ray observations of
the Crab Nebula with {\it Chandra}, whose spatial resolution of
$0.^{\!\!\prime \prime}5$ is comparable to that of ground-based
optical telescopes. We adopt a distance of 2 kpc to the Crab Nebula
throughout this paper.

\vspace{-0.3cm}
\section{Observation}
\vspace{-0.2cm}

The Crab Nebula was observed with ACIS-S3 (the back-illuminated CCD
chip) eight times, once every three weeks from 2000 November 3 to 2001
April 6. These observations were coordinated with the {\it Hubble
Space Telescope} (HST) (Hester et al. 2002). Each observation has
approximately 2.6 ksec of effective exposure time. We employed a 0.2
sec frame time to reduce pileup and a restricted window to reduce
dead-time. The window size is only slightly larger than the X-ray
extent of the Crab Nebula (see fig~\ref{fig:2ndimage}). All of the
images shown were made using the 0.2--10 keV band.

\vspace{-0.3cm}
\section{Result and Discussion}
\vspace{-0.2cm}

\begin{figure}[t]
\begin{center}
\hspace{0cm}
\psfig{file=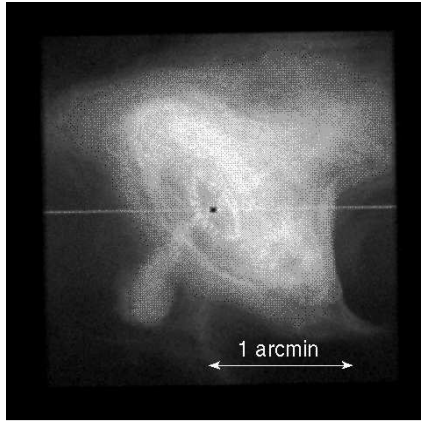,width=5.5cm}
\caption{{\it Chandra} ACIS-S image of the Crab Nebula observed on 2000
November 25 (2nd observation), displayed with a square root
brightness scale to enhance the faint structures. Due to heavy pileup
which results in on-board event rejection, there is a hole where no
events are detected at the position of the pulsar. The narrow line
through the pulsar is the so-called trailed image of the pulsars.}
\label{fig:2ndimage}
\end{center}
\vspace{-0.5cm}
\end{figure}

Figure~\ref{fig:2ndimage} shows one of eight images of the Crab
Nebula. It clearly shows an axisymmetric structure about the polar jet
with the torus and the inner ring resolved in an early {\it Chandra}
observation (Weisskopf et al. 2000). Fine fibrous structures are also
resolved at periphery of such large-scale structures. They show clear
correlation with a optical polarization measurement (Hickson \& van
den Bergh 1990), indicating that they trace local magnetic field structures.

\begin{figure}[t]
\begin{center}
\hspace{0cm}
\psfig{file=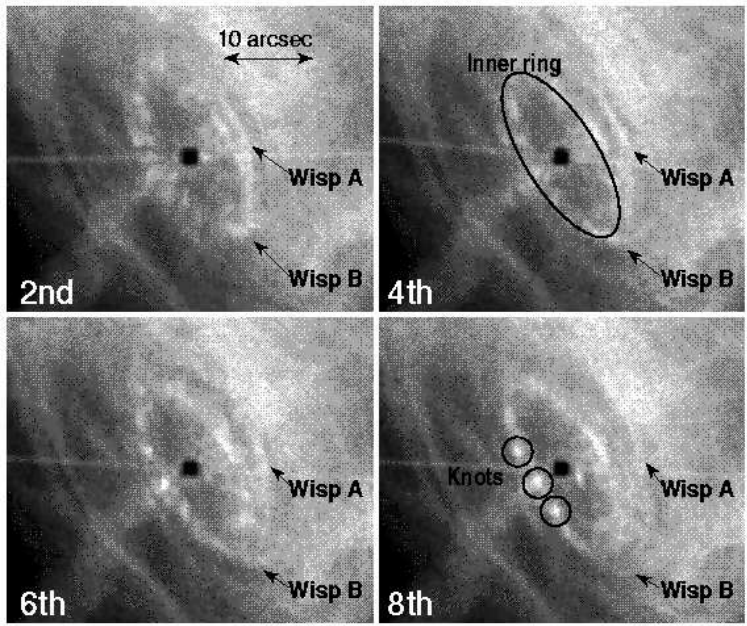,width=10cm}
\caption{Expanded images of the 2nd, 4th, 6th, and 8th observations, spaced
6 weeks apart. They demonstrate that two wisps are breaking off from
the inner ring and that the inner ring varies.}
\label{fig:2468image}
\end{center}
\vspace{-0.6cm}
\end{figure}

Figure~\ref{fig:2468image} shows a series of the observations. Two
wisps moving outward are detected through all of eight observations.
They can also be seen in simultaneous HST observations (Hester et al.
2002). Here, we denote them as ``wisp A'' and ``wisp B''. They appear
to break off from the inner ring. With respect to the inner ring's
elliptical shape, the shapes of the wisps look warped due to the time
delay of light travel (Hester et al. 2002). We measured the speed of
the wisps, assuming that they are moving in the equatorial plane which
includes the inner ring, and taking the inclination of the plane (the
inclination angle was derived assuming the inner ring is circular) and
the time delays into account. In spite of the difference in the
directions and the birth times, the speeds of the two wisps are almost
same, $\sim$0.43{\it c}. Their similarity indicates the existence of
the continuous isotropic pulsar wind in the equatorial plane.

The inner ring is also variable, but unlike the wisps, it preserves
its overall ring-like shape and relative position with respect to the
pulsar. The ring never forms a continuous loop, but appears
intermittent and mostly consists of knot-like features. Among them,
three knots, which lie along the southeast portion of the ring and are
symmetric about the axis of the jet, gradually brighten by factor
$\sim$1.5 within our observational period of 6 months. However, we
note that these knots were bright enough to be detected 1 year before
our 1st observation (Weisskopf et al. 2000). Additionally, some
blob-like features appear to move outward along the jet, with a speed
of $\sim$0.34{\it c}.

\begin{figure}[t]
\begin{center}
\hspace{0cm}
\psfig{file=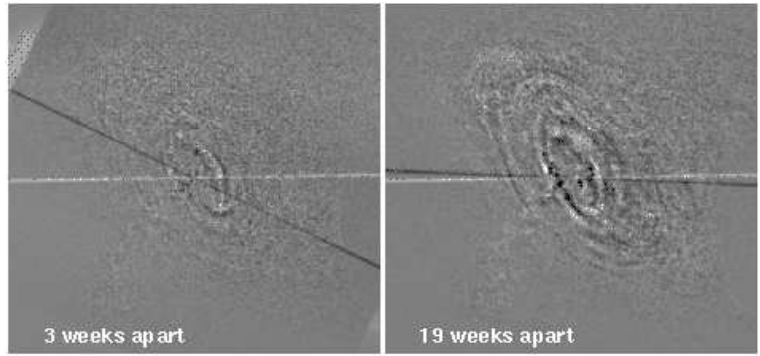,width=10cm}
\caption{Difference images of the 2nd$-$3rd (left), and the 2nd$-$8th (right) 
observations. Features which are strong in the earlier observation
appear as white in the images.  Longer time scale variations than that
of the wisps can be seen in the torus.}
\label{fig:diff}
\end{center}
\vspace{-0.6cm}
\end{figure}

Temporal variations can also be seen in the torus.
Figure~\ref{fig:diff} shows the difference images of the 2nd$-$3rd and
the 2nd$-$8th observations. Although the variation is not strongly
pronounced in the 3-week difference image, except for the wisps,
differences within the torus and along its boundary are quite
substantial over a duration of 19 weeks. The torus seems to be
expanding at 0.1--0.2{\it c}. Due to the small displacement and the
ambiguous boundary, relatively large uncertainties remain. The fact
that the angular extent along the major axis measured with {\it
Chandra} agrees well with those measured with {\it Einstein}
(Brinkmann et al. 1985), {\it ROSAT} (Hester et al. 1995;
Greiveldinger \& Aschenbach 1999), and even from lunar occultation 25
years ago (Aschenbach \& Brinkmann 1975) suggests that the torus is
stable on tens of arcsecond scales over decades, but varies on
arcsecond scales on several months. The northwestern region and the
end of the jet do not exhibit any strong variations.

Pavlov et al. (2001) discovered that the Vela PWN also shows temporal
variability. Thus, it is indicated that the dynamic behavior is a
common feature of PWNe.

These X-ray observational pictures generally match the canonical
scenario that the PWN is confined by the supernova ejecta. The
relativistic winds from the pulsar are continuously transported in the
equatorial plane and accumulate at the torus, outside of which the
optical filaments -- ejecta -- can be seen.

\end{document}